# First-principle study on the electronic and optical properties of cavities occupancy of SI methane hydrates

Zhao Wang[1], Liang Yang[1], Rong Deng[1], Zejin Yang[2]*

[1]*College of Materials Science and Chemical Engineering, Hainan University, Haikou 570228, China*

[2]*School of Science, Zhejiang University of technology, Hangzhou 310023, China*

*zejinyang@zjut.edu.cn

**Abstract**

The electronic density of state and optical properties of the SI methane hydrates with three different configurations are calculated by first-principle, including (abbreviated as cI, similarly hereinafter), each cage is fully filled with only one methane molecule; (cII), only one of the small cage lacks its unique methane molecule; (cIII), only one large cage lacks its unique methane molecule. The results show that the hydrate is most stable in the cI due to its perfect structure, the cII is less stable, whereas the cIII is least stable owing to its larger structural distortion after the loss of methane. On the contrary, loss of a methane from cII causes negligible deformation. The electronic density of states and the energy band gap of the cII is almost the same with those of cI, differing obviously with those of cIII. Methane hydrate is only responsive to light in the ultraviolet region, revealing their similar properties, regardless of their structural discrepancies, or/and the different ratios of water and methane molecules, 46/8=5.75 versus 46/7=6.57. Our calculations demonstrate that the lack of one methane in the cII causes negligible influence to the lattice structure and therefore to the electronic and optical properties in comparison with the cI, whereas the lack of one methane in the cIII can cause detectable changes. These results might provide valuable reference to the industrial exploration.

**INTRODUCTION**

Since gas hydrate plays an important role in potential energy, global climate change, submarine geological environment, and natural gas pipeline transportation, it has been an important issue in recent years[1-6]. Although natural gas hydrates are widely found in seabed sediments and terrestrial frozen soils, no commercially viable mining methods are available till now[7-9]. Over the last several decades, despite significant knowledge has been gained regarding its diverse properties, many aspects still remain puzzling or unknown[10-12].

Hydrates have three different kinds of crystal structures, consisting of cubic structure I (SI), cubic structure II (SII), or hexagonal structure H (SH), respectively. Ideally, all cavities are filled with one methane molecule. For ideal hydrates, the hydration number of the structure I (the water molecule number per guest molecule) is 5.75 (46/8). Amadeu K. Sum *et al* [13] measured cage occupancy and hydration number for $CH_4$ hydrate at different equilibrium conditions (temperature and pressure) and found about 90% occupation in the small cavities and almost fully filled in the large cavities. Satoshi Takeya *et al* obtained occupation 99% in the large cages and 69% in the small cages for $CO_2$ per unit cell by the powder X-ray diffraction technique, respectively, slightly differing with previous values from single crystal analysis [14]. The value of the hydration number ranges from about 5.8 to 6.3 in above literatures. The theoretical prediction for cage occupancy and hydration number have been reported based on the statistical van der Waals–Platteeuw model together with *ab initio* calculations and molecular dynamics method [15-17].

In the detection of hydrate reserves in seabed sediments, seismic velocity techniques as well as bottom simulating reflections are the main remote means [18-19]. The methods of drilling and coring are able to refine and estimate the reservoir hydrate content. Logging tools such as caliper, gamma ray, density, resistivity, and neutron porosity are used to determine the hydrate depth and the concentration [20-23]. With Time Domain Reflectometry, the gas hydrate amount is measured based on the bulk dielectric properties [24]. These methods generally require some parameters such as dielectric constant. Unfortunately, these parameters are relatively lacking and controversial. In some literatures, the dielectric constant of hydrates is substituted by the dielectric constant of ice [25].

In this work, we use a first-principle method to calculate the electronic and optical properties of hydrates in order to obtain the important parameters such as dielectric coefficients. First principle method is widely used tool to study the solid state physics till now[26-48]. Since hydrates are non-stoichiometric substances, it is required to consider the effects of different cage occupancy for these parameters. In addition to the calculation of the cI (the correspondent hydration number is 46/8 = 5.75), we also performed the same calculation for the cII and cIII (hydration number 46/7 = 6.57, the cage occupancy rate is 87.5%.). We intend to reveal the effect of cage occupancy to the optical and electrical properties.

**COMPUTATIONAL SETUP**

In this paper, the hydrate coordinates are determined according to the available literature[49]. The first-principle calculations were performed by VASP [50,51] and the projector-augmented wave function. The energy cutoff for plane-wave expansion was set to 600 eV. The first Brillouin zone sampling was chosen as 2×2×2 $k$-point mesh per unit cell . The optimized forces on all atoms were converged to below 0.02 eV/Å. Electronic relaxations were converged to within $1\times10^{-8}$ eV. The optical properties were calculated by CASTEP code[52,53], the detailed calculation process can be found in its manual, for simplicity, we ignore the concrete description.

**RESULTS AND DISCUSSION**

We carry out first-principles calculations for three different types of SI-hydrate structures, including: (I), each cage is filled with methane molecules in the unit cell (46 water molecules, including 6 large cages, $5^{12}6^{2}$, 2 small cages, $5^{12}$), with a hydration number of 46/8; (II), only one of the small cage lacks its methane molecule, all of the other cages are still fully filled with respective methane molecules, with a hydration number of 46/7; (III), only one large cage lacks its one methane molecule and all of the other cages are still fully filled with respective methane molecules, with a hydration number of 46/7, respectively.

The ideal structure is that each cavity is fully filled with one methane molecule, as shown in Figure 1. The relaxed cell parameters are $a=b=c=11.62$ Å, with a cell volume of 1568.98 Å$^3$, $\alpha=\beta=\gamma=90.0°$.

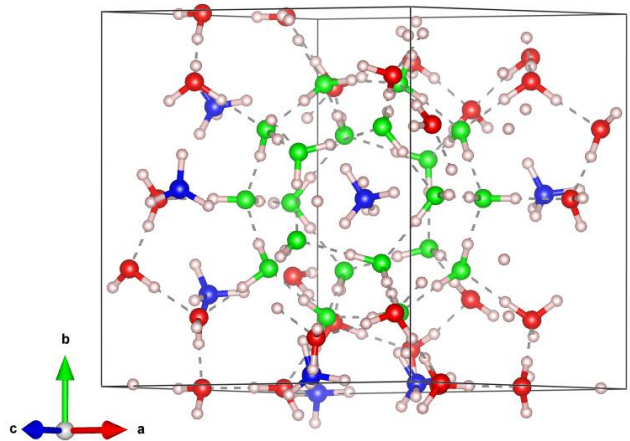

Figure 1. The hydrate unit cell structure, each cavity is fully filled by one methane molecule, in which a small cage is highlighted by green color, the other water molecules are red color.

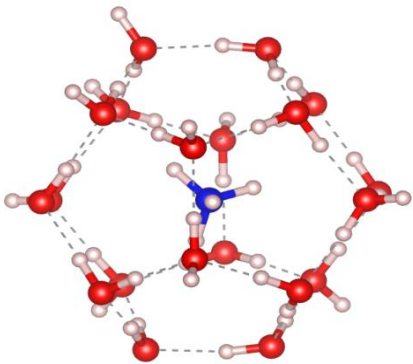
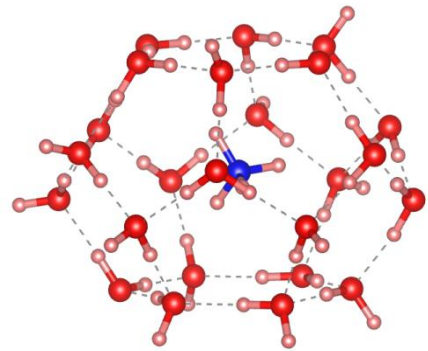

(a)                          (b)

Figure 2, The two different building blocks of the hydrate structure, consisting of one small cage (left) and one large cage (right), the nearly spherical shape of small cage is clearly seen.

The relaxed cell parameters are shown in Table 1, from which we can see that the absence of one methane molecule from the small cage failed to induce evidently changes relative to that of perfect structure, whereas the structure distorted obviously in the case of large cage. The large deformation might relate to the large hollow cavity, which corresponds to small resistance to the neighbor squeezing interaction.

Table 1. cell parameters

|  | $a$ (Å) | $b$ (Å) | $c$ (Å) | $\alpha$ | $\beta$ | $\gamma$ |
|---|---|---|---|---|---|---|
| full cage | 11.62 | 11.62 | 11.62 | 90.00° | 90.00° | 90.00° |
| small cage lack | 11.61 | 11.62 | 11.59 | 90.12° | 89.89° | 90.05° |
| large cage lack | 10.85 | 12.72 | 11.26 | 86.86° | 96.11° | 94.65° |

The perfect lattice energy is -876.7 eV, lower than those of defective structures, -852.7 and -845.9 eV for the methane lacks from small and large cages, respectively, it is also clear that the hydrate is more stable when methane lack from the small cage than that of large cage, which might be one possible explanation to the experimental observation of the higher occupancy rate of the small cage other than the large cage, still consistent with the experimental observation that the small cage is formed first and the large cage subsequently[54].

The fully filled cage (cI) has a distance of 3.77 Å between the C and O atoms in the small cage, slightly smaller than that of experimental data, 3.95 Å[55] and the length of pentagon is 2.7 Å, both which are almost identical with those of small cage lack structures. The experimental lattice parameters[55] are also slightly larger the calculated value, 11.725-11.932Å. Moreover, all of the nearest neighbors of the small cage are large cages, such higher symmetric environment should be one of the origins of the nearly unchanged lattice structure once its methane is removed. The undistorted structure should also be attributed partially to its nearly spherical shape building blocks in the unit cell. However, the inequivalent nearest neighbors of the large cage, comprising of large and small cages simultaneously, results in its lower symmetry in chemistry, which is quite difficult to keep its original shape. Owing to the giant deformation of the lattice, it is therefore not meaningful to compare the inter-atomic distance with the ideal lattice. The elongated lattice parameter $b$ could be illustrated by its shorter vertical distance of two hexagonal planes than that of the diameter of its perpendicular plane (nearly circle), that is, totally depending on the building block orientations of the large cage in the lattice, as is shown in Figure 2 (b), vice versa to the case of the shortened $a/c$.

1. **The electronic density of states**

We calculated the electronic density of state (DOS) for the three configurations, as shown in Figure 3. Owing to the large structural change when one methane is removed from a large cage, the degree of degeneracy state is decreased and the energy level is broadened, and the resolution or the full width at half maximum (FMHM) in the valence band profile is increased. Loss a methane from a small cage failed to induce significant structural distortion, as is also the feature in the DOS in comparison with the perfect structure. Their energy gaps between the bottom of conduction and the top of valence bands, 5.27, 5.23, 3.81 eV for cI, cII, cIII, respectively.

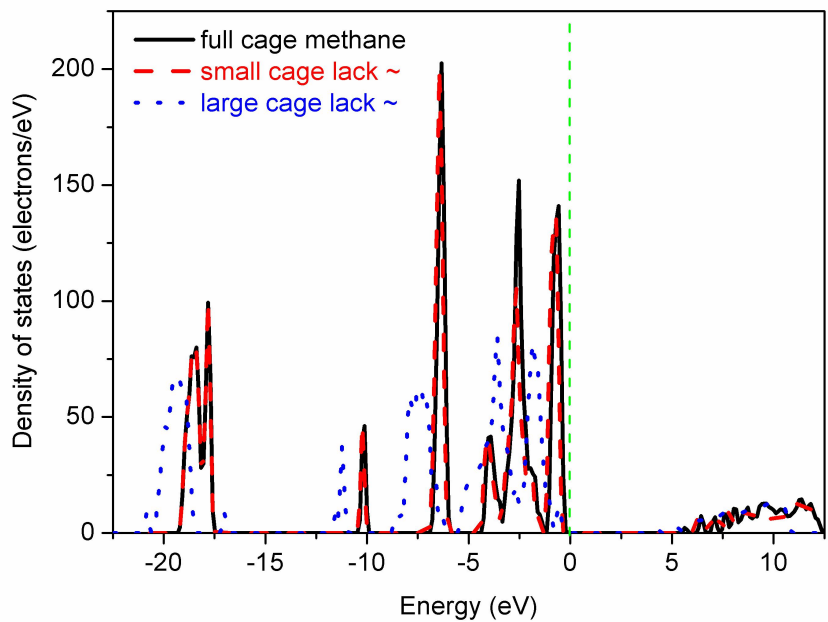

Figure 3. The electronic density of state for the three configurations.

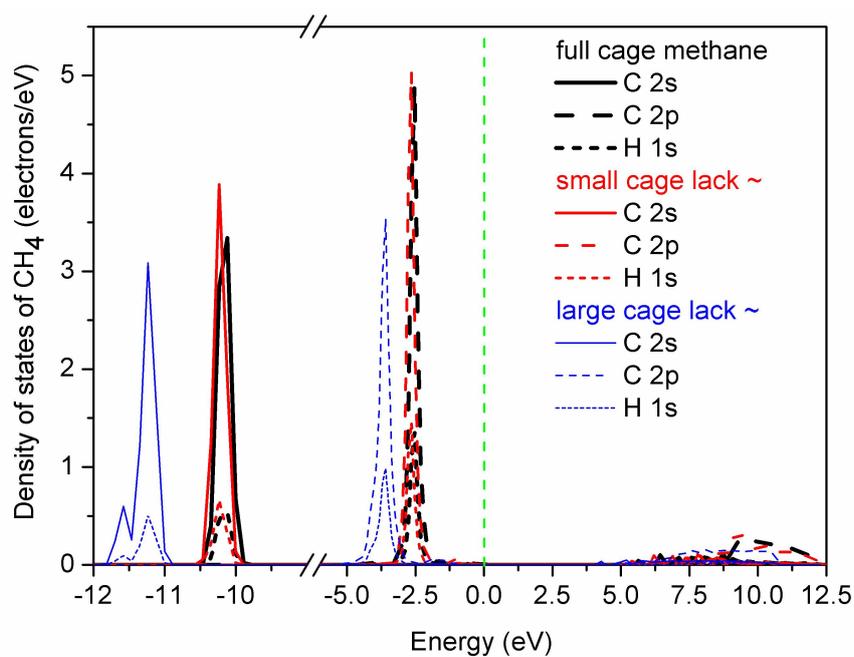

Figure 4. The electronic density of state in three configurations for methane molecule.

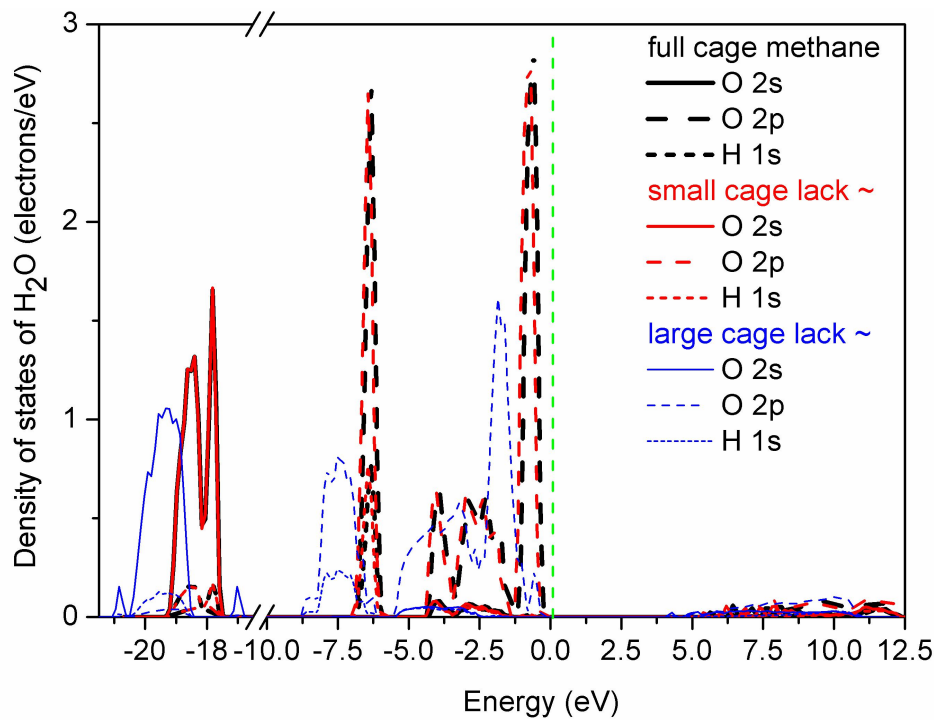

Figure 5. The electronic density of state in three configurations for water molecule.

In order to compare the changes of atomic DOS during the structural change, the partial density of state (PDOS) are calculated and shown in Figure 4, 5, in which the giant energy level shift towards lower energy side, ~1 eV, including methane and water in cIII, meaning the giant internal stress. However, some O 2p state still keeps almost unchanged in cIII in Figure 5, denoting some unchanged water molecules site. The decreased intensity means the strong delocalized distribution features. Note that the data are the average value for all of same atom kind, otherwise we need to plot too many figures as the atomic PDOS shows strong site dependence. Moreover, the shift of energy level towards lower energy side in methane and water decreased the hydrogen bonding interaction and therefore decreased the structural stability. The more delocalized distributions of O 2p and H 1s states reflect the non-uniform hydrogen-bonding characteristics, illustrating the more disorder nature of the water site.

Due to the negligible distortion of cII, the lattice still remains highly symmetric feature, thus the atomic orbitals are still strongly degenerate and remains unchanged.

2. Dielectric constant

The real and imaginary components of the dielectric coefficient at different frequencies for the three configurations are shown in Figure 5. Generally, all of them display similar profiles besides the static values, careful analysis one can detects the evident change from cIII, whether in real or imaginary parts, respectively. Both the real and the imaginary parts of the dielectric coefficient in cI and cII are nearly identical. The relative intensity is reduced and behaved double peaks at frequency range of ~5-7 eV in the methane absence of large cage, which results in the profile more flat and enhances the detectability. The static dielectric constants are 1.7 for the three configurations, far smaller than some literature reports ~58[1,56], resulting mainly from the large contribution of proton disorder. The dielectric constant appears to be insensitive to the trapped methane molecules, depending mainly on the cage-shaped structure of the water molecule based on the general profile invariance nature. The proton disorder play a dominant role for the dielectric constant of ice or hydrate, and its calculation formula is [56-60].

$$\varepsilon = \varepsilon_0 + (\frac{4\pi}{3Vk_BT})(\langle M^2 \rangle - \langle M \rangle^2) \qquad (1)$$

Where $\varepsilon_0$ is the optical dielectric constant and M is the electric dipole moment. The optical dielectric constant $\varepsilon_0$ is set equal to 1.592[57]. The high dielectric constant is only for pure ice or hydrates. The dielectric constant is about 3[59] if the ice contains impurities,. Wright et al. report the hydrates dielectric constant relationship with the sample moisture content $\theta$[24].

$$K_a = 4.0556 + 0.1132\theta + 0.007869\theta^2 + 0.00002169\theta^3 \qquad (2)$$

Here, the apparent dielectric constant $K_a$ is the dielectric constant of the mixture other than the pure hydrate. The $K_a$ is 4.0556 when the water content is zero. Our calculated dielectric constant is only involved for the electrons, which is equivalent to ε₀, that is, the optical permittivity. Protons and their optical effects have been neglected. Obviously, the dielectric constant value of methane hydrate is still controversial. Since the protons disorder effects have been neglected, the results obtained here can be used as a reference to the value of $\varepsilon_0$ in (1).

3. **The other optical properties.**

The reflectivity index for the three configurations are shown in Figures 6. Similarly to the distribution of dielectric profiles, the main peak of reflectivity index are also located at around 5~12 eV.

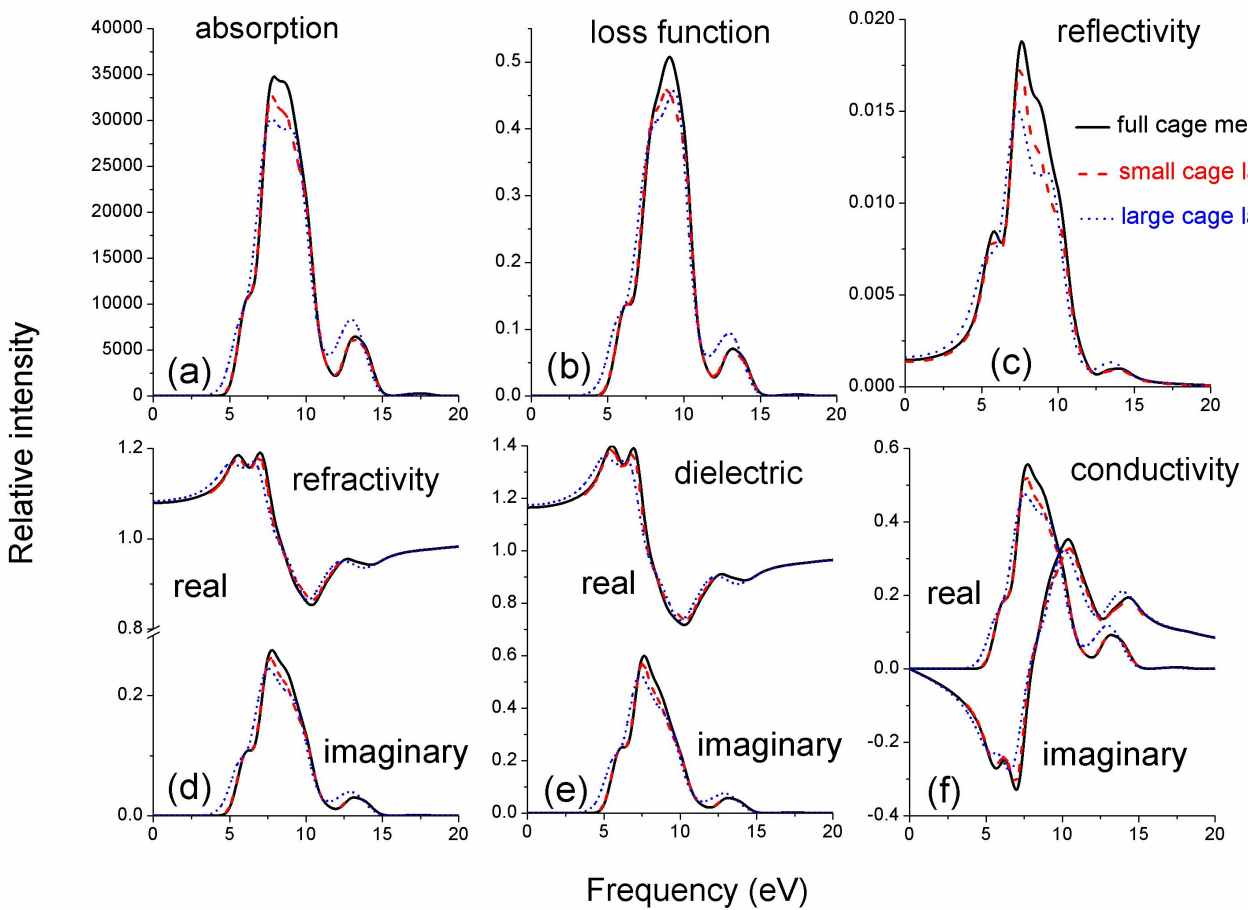

Figures 6. The absorption coefficient (cm⁻¹), refractive index, extinction spectrum, energy loss spectrum, dielectric constants, conductivity (1/fs) of three different configurations, arbitrary units are used for the unmarked quantities.

The absorption, refractive index, extinction spectrum and energy loss spectrum are shown in Figure 6. The reflected light of the gas hydrate is distributed at the ultraviolet region, which is true in the other optical properties. The absorption peak reflects the transition of water 2p electrons. A wide platform of peak maximum totally agree with the complicated O 2p and H 1s DOS in cIII. Furthmore, the maximum absorption intensity is also decreased, originating from the decreased PDOS of water. The refractivity index presents similar variations with those of dielectric constant. In fact, the intensities of these four functions show strong structure dependence, that is, the general profiles remain almost unchanged. The intensity of conductivity index is substantially decreased and also present similar profile with that of absorption coefficient, agreeing with the fact of large energy level shift in PDOS of methane. The most sensitivity region locates at 6~8 eV for all these optical properties. It is therefore reasonable to infer that these properties are dominated by the host molecule cage framework structure. The guest molecule does not play decisive role in the optical properties.

**CONCLUSIONS**

The lattice structure is more stable when a methane molecule is removed from a small cage than that of large cage because the small cage is nearly spherical symmetry and its equivalent chemical environment. Loss a methane molecule from a small cage induce subtle structural distortion, electronic density of states shift, energy gap reduction, optical related properties response. However, such electronic and optical properties change obviously owing to the larger structural deformation once loss a methane from a large cage. The present investigation also provides valuable insight to other defective cage framework structure with similar variations. The obtained dielectric function shows nearly same profiles with that of refractivity index. The absorption coefficient, loss function, reflectivity index, and the real part of the conductivity index also present similar response. The sensitivity response range is about 5-15 eV for all of the optical properties. The current investigations demonstrate the correlation of the structural change of optical and electrical properties with the degree of structural distortion. The negligible structural and electronic change of methane lack from small cage demonstrates the extreme stability of the water framework, independing on the trapped guest molecules, occupancy or not, which is the true origin of the substance storage capacity.

**ACKNOWLEDGMENT**

Wang Zhao acknowledges the financial support from the Hainan province key research and development plan projects (No. ZDYF2017098), Zejin Yang acknowledges the Natural Science Foundation of Zhejiang Province, China (Grant No. LY18E010007).